\documentclass[sigconf]{acmart}

\usepackage{tabularx} 
\usepackage{multirow}

\AtBeginDocument{%
  \providecommand\BibTeX{{%
    \normalfont B\kern-0.5em{\scshape i\kern-0.25em b}\kern-0.8em\TeX}}}

\copyrightyear{2021} 
\acmYear{2021} 
\setcopyright{acmlicensed}\acmConference[CHI '21]{CHI Conference on Human Factors in Computing Systems}{May 8--13, 2021}{Yokohama, Japan}
\acmBooktitle{CHI Conference on Human Factors in Computing Systems (CHI '21), May 8--13, 2021, Yokohama, Japan}
\acmPrice{15.00}
\acmDOI{10.1145/3411764.3445515}
\acmISBN{978-1-4503-8096-6/21/05}

\begin{document}

%%
%% The "title" command has an optional parameter,
%% allowing the author to define a "short title" to be used in page headers.
\title[Streaming VR Games to the Broad Audience]{Streaming VR Games to the Broad Audience: A Comparison of the First-Person and Third-Person Perspectives}

%%
%% The "author" command and its associated commands are used to define
%% the authors and their affiliations.
%% Of note is the shared affiliation of the first two authors, and the
%% "authornote" and "authornotemark" commands
%% used to denote shared contribution to the research.
\author{Katharina Emmerich}
\email{katharina.emmerich@uni-due.de}
\affiliation{
  \institution{University of Duisburg-Essen}
  \city{Duisburg}
  \country{Germany}
}

\author{Andrey Krekhov}
\email{andrey.krekhov@uni-due.de}
\affiliation{
  \institution{University of Duisburg-Essen}
  \city{Duisburg}
  \country{Germany}
}

\author{Sebastian Cmentowski}
\email{sebastian.cmentowski@uni-due.de}
\affiliation{
  \institution{University of Duisburg-Essen}
  \city{Duisburg}
  \country{Germany}
}

\author{Jens Kr{\"u}ger}
\email{jens.krueger@uni-due.de}
\affiliation{
  \institution{University of Duisburg-Essen}
  \city{Duisburg}
  \country{Germany}
}

%%
%% By default, the full list of authors will be used in the page
%% headers. Often, this list is too long, and will overlap
%% other information printed in the page headers. This command allows
%% the author to define a more concise list
%% of authors' names for this purpose.
\renewcommand{\shortauthors}{Emmerich and Krekhov, et al.}

%%
%% The abstract is a short summary of the work to be presented in the
%% article.
\begin{abstract} 
The spectatorship experience for virtual reality (VR) games differs strongly from its non-VR precursor. When watching non-VR games on platforms such as Twitch, spectators just see what the player sees, as the physical interaction is mostly unimportant for the overall impression. In VR, the immersive full-body interaction is a crucial part of the player experience. Hence, content creators, such as streamers, often rely on green screens or similar solutions to offer a mixed-reality third-person view to disclose their full-body actions. 
Our work compares the most popular realizations of the first-person and the third-person perspective in an online survey (\textit{N}~=~217) with three different VR games.
Contrary to the current trend to stream in third-person, our key result is that most viewers prefer the first-person version, which they attribute mostly to the better focus on in-game actions and higher involvement. Based on the study insights, we provide design recommendations for both perspectives.
  
\end{abstract}

%%
%% The code below is generated by the tool at http://dl.acm.org/ccs.cfm.
%% Please copy and paste the code instead of the example below.
%%


%%
%% Keywords. The author(s) should pick words that accurately describe
%% the work being presented. Separate the keywords with commas.

%% A "teaser" image appears between the author and affiliation
%% information and the body of the document, and typically spans the
%% page.
\begin{teaserfigure}
 \includegraphics[width=\textwidth]{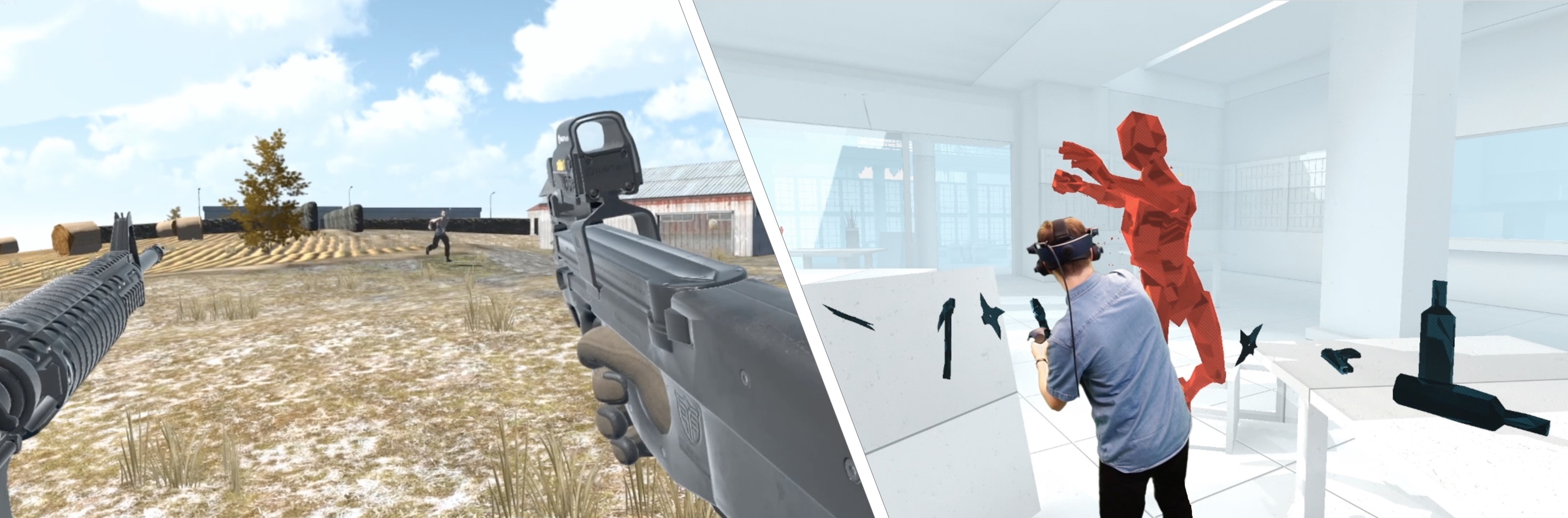}
  \caption{Our research compares two common perspectives used to deliver VR gaming content: using the player's view as a first-person perspective (left) versus a third-person mixed-reality view blending the player into the virtual world (right).}
  \label{fig:teaser}
  \Description[Teaser image contrasting the first-person and the third-person perspective of VR game streams]{This figures is the teaser image and shows two screenshots next to each other, which show two different video perspectives. The left screenshot is from the game Stand Out: VR Battle Royale and shows the first-person perspective. The right screenshot is from the game Superhot: VR and shows a mixed-reality third-person view, which shows the player blended into the game world.}
\end{teaserfigure}

%%
%% This command processes the author and affiliation and title
%% information and builds the first part of the formatted document.
\maketitle

\section{Introduction}
Ever since the establishment of live streaming platforms such as Twitch~\cite{twitch}, watching other play became a popular spare-time activity across all ages. The digital audience tunes in for various reasons, be it to follow an esports tournament or to check out a newly released, trending game. This curiosity also applies to recent virtual reality (VR) titles (cf. \textit{Half-Life: Alyx}~\cite{Alyx}), forcing streamers to adapt their content creation and delivery pipeline to the specifics of VR.
	
The main difference is that VR games heavily rely on the high degree of immersion provided by such stereoscopic setups. The player experiences a feeling of being in the virtual world, which is usually achieved by the head-orientation-dependent view in combination with realistic full-body interactions. Clearly, that impression cannot be easily transferred to the audience, because most spectators utilize 2D displays, e.g., mobile devices or PC/TV screens. Hence, streamers seek for viable non-VR workarounds to deliver the immersive VR gaming content.

One prominent way to transport the experienced presence to the audience is to provide a mixed-reality view of the player/streamer. By switching to a third-person perspective, the player blends into the surrounding virtual environment. The spectators can see the player’s full-body movements and interactions in context, enabling a better understanding of the actual gameplay experience (see Figure~\ref{fig:teaser}).

On the other hand, the traditional first-person perspective offers a unique advantage: seeing the game from the player’s perspective brings the spectators as close to the in-game action as possible. Due to the same point of view, the spectators obtain identical visual information. This similar visual perception of the virtual world can potentially evoke the viewers’ feeling of playing on their own.

So, which perspective is best for the streaming of VR games? Given the aforementioned conceptual differences and the fact that both perspectives are widely used and accepted, we do not expect a definite answer to this question. The right choice of perspective seems to depend on different contextual factors, such as the type of the game or the purpose of watching. Hence, there is a need to study spectators' preferences and motives in different contexts to support the creation of compelling audience experiences.

With our work, we lay the foundations for the research of spectator experiences and perspectives in VR settings. The choice of perspective significantly frames the viewing experience. While researchers agree that immersion is an important part of interactive VR experiences, it remains unclear how important immersion is for spectators of VR players compared to other factors such as contextual understanding and player centricity.

Aiming towards a comprehensive VR streaming guideline, our work is the first to contribute relevant insights into spectators’ opinions. We present an online survey (\textit{N}~=~217), which covered three different VR games: \textit{Beat Saber}~\cite{BeatSaber}, \textit{Superhot VR}~\cite{Superhot}, and \textit{Stand Out: VR Battle Royale}~\cite{StandOut}. For each game, the spectators watched a first-person and a third-person video and finally shared their impressions. The so obtained results allow us to discuss each perspective’s particular strengths and weaknesses and formulate preliminary design considerations, which are meant to provide a starting point for VR content creators.

We have to understand how different perspectives (such as first-person and third-person) contribute to different demands of spectators to be able to make informed design choices. This research is not only relevant in the context of game streaming. It also applies to related VR setups including some kind of spectator. In particular, the choice of perspective is important in multi-user scenarios that combine VR and non-VR users. Example scenarios include VR training applications (surgical training, rehabilitation games) where the perspective choice is crucial for supervisors to be able to adequately monitor and evaluate the trainee's performance. Hence, apart from giving practical advice to VR streamers and content providers, our work creates the basis for more sophisticated choices of perspective and paves the way for future research on audience experiences in VR.

\section{Related Work}
 
Today, online video and streaming platforms such as YouTube~\cite{youtube} and Twitch~\cite{twitch} enable a globally distributed audience to watch their favorite players and games at any time. Former consumers can now easily produce user-generated content (UGC) and are challenging the traditional media~\cite{cha2007tube}. 
So-called \textit{Let's Play} videos have become increasingly popular~\cite{glas2015vicarious} and game live streaming has become a cultural phenomenon comparable to sports events~\cite{hamilton2014streaming, pires2015youtube, smith2013live}. 
Apart from casual gaming videos, competitive gaming events, commonly referred to as esports~\cite{hamari2017esports, rambusch2017pre}, are taking a growing share of the overall streaming landscape~\cite{cryan2014esports}.

Considering the overall popularity of game streaming, the motives and experiences of spectators have been of ongoing interest to the games user research community~\cite{kappen2014engaged, taylor2016play, wehbe2015towards}.
Instead of just focusing on the experience of the active player, 
a variety of work has broadened the research scope by explicitly investigating the spectator experience~\cite{carlsson2015designing, drucker2003spectator, frome2004helpless, maurer2015gaze, tekin2017ways} and the motivations for viewers to spend their free time watching others play~\cite{downs2014audience, hamilton2014streaming, kaytoue2012watch}. The spectator's experience is influenced by the game content, but also by the spectator interface~\cite{reeves2005designing}, that is the available information and the perspective from which they view both the game and the player.
Moreover, spectators---no matter if co-located or mediated---are part of the social play setting and often engage in some form of interaction with the player, which further shapes their experience~\cite{Emmerich.2019, tekin2017ways, voida2009wii, hamilton2014streaming}. 

This social interaction is also a strong motivator for watching game streams~\cite{hamilton2014streaming}. Other vital factors are enjoyment, information seeking, and distraction~\cite{cheung2011starcraft, hamilton2014streaming, kaytoue2012watch}.  
For esports, research has revealed two additional motivators: the general competitive atmosphere and the opportunity to share emotional connections~\cite{hamari2017esports, lee2011comparison, shaw2014sport, weiss2013virtual, weiss2011fulfilling}. Hence, the reasons why spectators watch Let's Play videos and game live streams seem manifold. 
A commonly used framework to assess the motivation behind media usage is the \textit{uses and gratification} (UG) model~\cite{katz1973uses, katz1973use, rubin2009uses, ruggiero2000uses}. UG is based on the assumption that users actively choose certain media with the motivation to achieve a particular gratification. The available media has to compete constantly with other sources of gratification, and personal reasoning is considered individually for every user. UG typically classifies the user needs into the categories \textit{cognitive, affective, personal integrative, social integrative, and tension release}. In an empirical study, Sjöblom et al.~\cite{sjoblom2017people} revealed that all five classes of gratification are associated with the motivations of twitch users watching game live streams.

The previous research on game videos and streams is mainly based on the footage of common, non-VR games.
VR games have just recently gained a foothold in the consumer market. New hardware and the release of sophisticated AAA-games, such as \textit{Half-Life: Alyx}~\cite{Alyx} and \textit{Asgard's Wrath}~\cite{AsgardsWrath}, have led to regular media coverage and an increased interest of the broad gamer community. 
In contrast to non-VR games, the special setup of VR games including head-mounted displays (HMDs) and movement tracking makes it more challenging to convey the entire immersive experience to spectators. In particular, the player's body movements used to control the game become an integral part of gameplay.  
Research dealing with the audience of VR games remains sparse and is mainly focused on local spectatorship~\cite{gugenheimer2017sharevr, hartmannRealityCheck, jones2014roomalive, welsfordAsymmetric, krekhov2020silhouette}.
Hence, the questions remains how the game experience can best be delivered in videos and streams to a broad online audience. 

One possibility is to use the same approach as with non-VR games: players directly broadcast the game view that is displayed on the HMD, so that the audience sees the same as the player. This first-person perspective ensures that the spectator's focus always matches the player's current focus and that the spectators see the game world as if they were playing themselves. Research on different player perspectives indicates that a first-person view makes it easier for players to focus on the action and provides advantages to immersion~\cite{denisova2015first, voorhees2012guns}. These effects might also apply to the spectator's perspective.

On the other hand, a first-person streaming perspective does not show the player's bodily interaction with the VR game. This might impede a full understanding of what is happening, as the manipulations conducted by the player are partly hidden from the spectator and only the effects in the game are revealed~\cite{reeves2005designing}. Tekin and Reeves~\cite{tekin2017ways} point out that seeing the game on screen and at the same time seeing the player's bodily actions---resulting in a ``dual vision''---are important parts of spectating.  
Therefore, many VR game streamers follow a different approach by providing a third-person perspective: they use a green screen and external cameras to blend themselves into the virtual world (similar to the original trailer of the HTC Vive~\cite{viveTrailer}). This mixed-reality perspective enables spectators to see the game world and the player at the same time, and might thus enhance their experience. At the same time, this approach creates a mismatch between the spectator's view and the player's view. Consequently, the third-person perspective underlines the difference between player and spectator and shifts the focus of spectating from events in the game world to the player.

As both perspectives seem to have advantages and shortcomings, the question arises how spectators experience and evaluate the different views and which perspective is superior in certain contexts. 
To the current date, there is no other work investigating this open research question. While there are some related studies on the use of different user perspectives in VR environments~\cite{cmentowski2019outstanding, gorisse2017first, salamin2006benefits, slater2010first}, these are not directly applicable to the experience of spectators.

\begin{figure}
\centering
\includegraphics[width=\columnwidth]{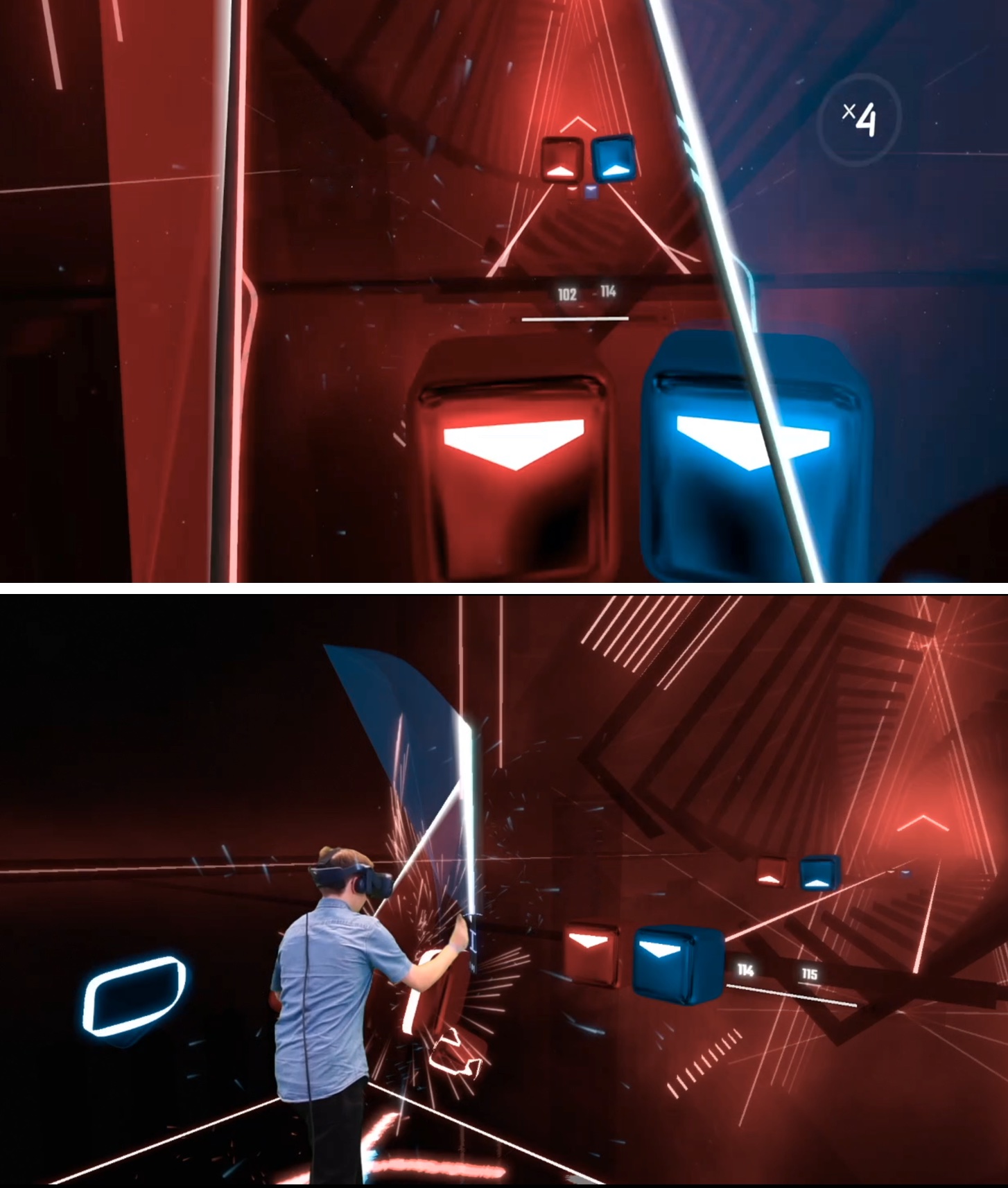}
\caption{\textit{Beat Saber}~\cite{BeatSaber} is one of the three games used in the online survey. Participants compared two perspectives: first-person (top) and third-person view (bottom).}
\label{fig:beatsaber}
\Description[Comparison of the first-person and the third-person perspective for the game Beat Saber]{Two in-game screenshots of the game Beat Saber show the two video perspectives used in the survey. The upper screenshot shows the game world from the first-person perspective, depicting colored blocks approaching the camera. The screenshot below shows the same game scene from a different angle and additionally includes a recording of the real player who is blended into the game environment, swinging two lightsabers.}
\end{figure}

\begin{figure}
\centering
\includegraphics[width=\columnwidth]{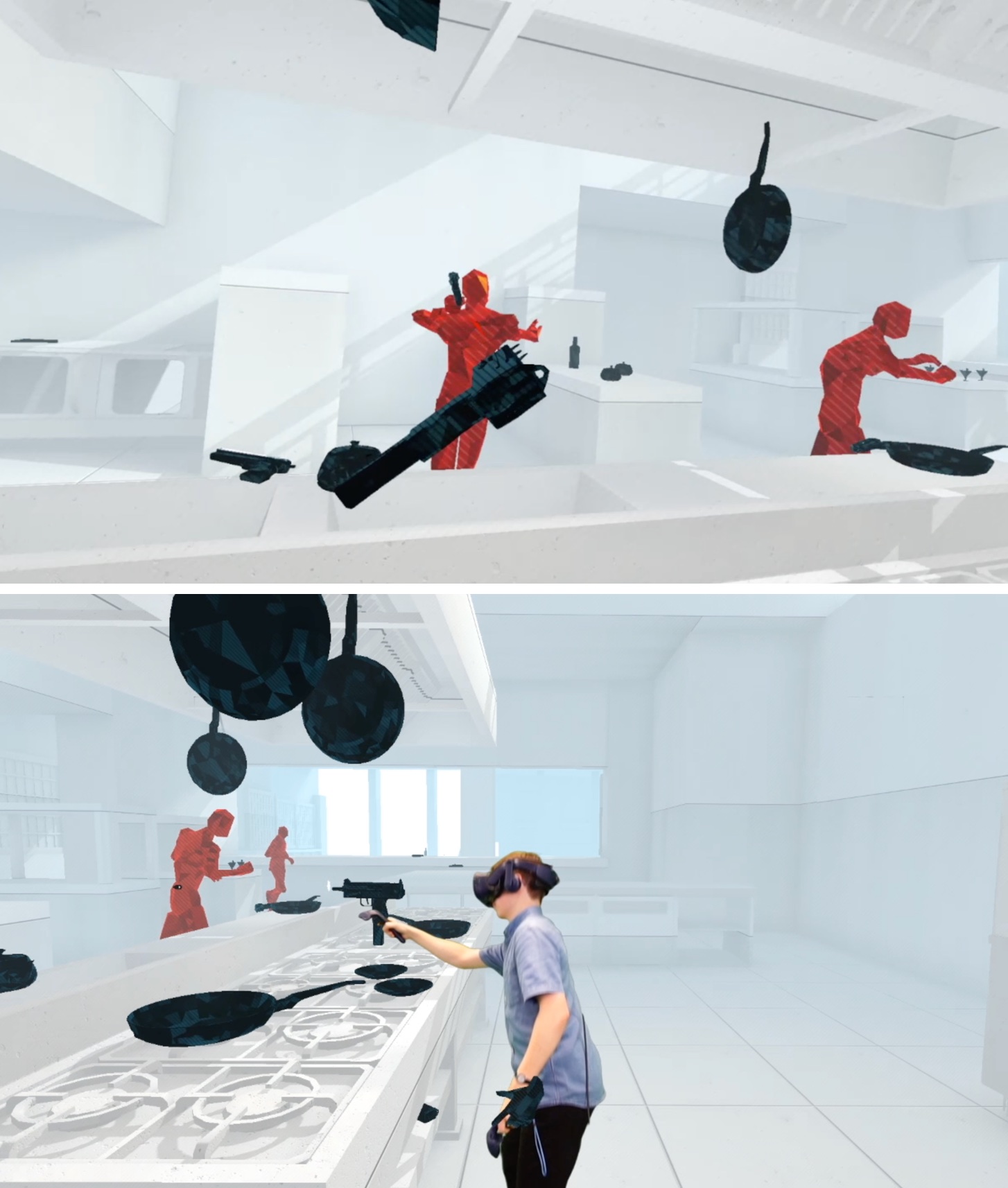}
\caption{\textit{Superhot VR}~\cite{Superhot} is one of the three games used in the online survey. Participants compared two perspectives: first-person (top) and third-person view (bottom).}
\label{fig:superhot}
\Description[Comparison of the first-person and the third-person perspective for the game Superhot VR]{Two in-game screenshots of the game Superhot VR show the two video perspectives used in the survey. The upper screenshot shows the game world from the first-person perspective, depicting a kitchen room with two enemies and several objects such as weapons and pans. The screenshot below shows the same game scene from a different angle and additionally includes a recording of the real player who is blended into the game environment.}
\end{figure}

\section{Online Survey}
We conducted an online survey to assess spectators' preferences and opinions on the different perspectives in VR game videos.
More precisely, we compared the first-person perspective, which directly displays the in-game view of the player, with a mixed-reality third-person perspective, in which the player is captured and directly cut into the game world (see Figure \ref{fig:teaser}).
The goal of the study was to gain insights about the advantages and disadvantages of both approaches regarding different aspects of the viewing experience such as comprehensibility, entertainment, and involvement. Hence, our main research question is how spectators experience the two perspectives, which differences can be found, and which perspective is preferable in certain settings. 

\subsection{Selection of Three Exemplary VR Games}
We decided to compare the two perspectives using different commercial VR games, as viewers' preferences and experiences may also depend on certain characteristics of the game. 
Our game selection process was based on several criteria. First, the games should be popular and positively rated, to ensure that they provide an interesting experience and successfully make use of VR headsets. Second, the games had to support the software tool LIV~\cite{LIV}, which enabled us to create the mixed reality third-person perspective.
Finally, the games should represent different game genres, which feature different core mechanics and controls.
Following these criteria, we reviewed the rankings of current VR games on the online gaming platform Steam~\cite{steam} and analyzed viewer numbers on Twitch to identify popular games. 
We chose three games that match all criteria: \textit{Beat Saber}~\cite{BeatSaber}, \textit{Superhot VR}~\cite{Superhot}, and \textit{Stand Out: VR Battle Royale}~\cite{StandOut} (hereafter abbreviated as \textit{Stand Out}). All games had more than 25.000 peak viewers on Twitch and more than 1.000 mainly positive reviews on Steam, indicating their popularity.

\textit{Beat Saber} is a music-based VR game. The player chooses a song and then swings two colored lightsabers to cut blocks of the same color, which represent the beats of the music and quickly approach the player (see Figure~\ref{fig:beatsaber}).
Hence, the main focus of the game is on the quick gestural reaction of the player to the fast-paced blocks. There is a direct mapping between the player's hand movement and the movement of the lightsabers in the game. Apart from single steps to the side to avoid an obstacle wall, there is no locomotion needed. As the blocks always approach on fixed paths in front of the player, the view orientation in \textit{Beat Saber} is rather fixed. We chose this game due to its remarkably high popularity, and because in current \textit{Beat Saber} streams, both perspectives we want to compare (first person and third person mixed reality) are commonly used.

In \textit{Superhot VR}, the player has to complete short levels by destroying all enemies and dodging their attacks  (see Figure~\ref{fig:superhot}). For this purpose, the player can use various objects lying around, such as pistols and bottles. The unique twist of this game is that time progresses only at the speed in which the player moves. That means, if the player moves slowly, the enemies also move slowly, and vice versa. This way, the player has to consider every movement, resulting in rather slow-paced gameplay. Though the player can move in room-scale, the enemies are then approaching quickly. So in most levels, there is not much locomotion happening, and the focus is on the opponents. However, enemies are approaching from different sides, so that the perspective is not fixed in contrast to \textit{Beat Saber}.

\begin{figure}
\centering
\includegraphics[width=\columnwidth]{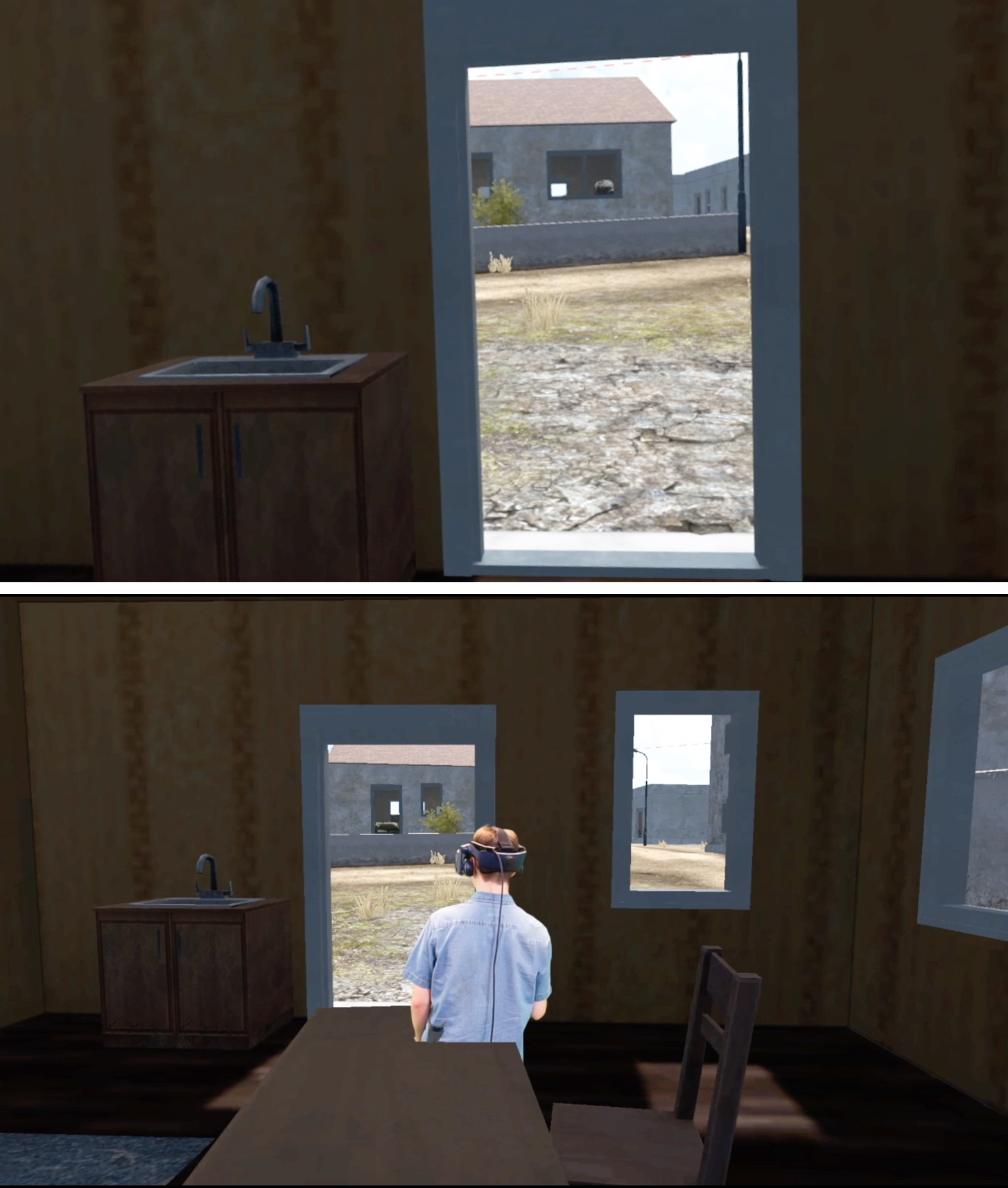}
\caption{\textit{Stand Out: VR Battle Royale}~\cite{StandOut} is one of the three games used in the survey. Participants compared two perspectives: first-person (top) and third-person view (bottom).}
\label{fig:standout}
\Description[Comparison of the first-person and the third-person perspective for the game Stand Out: VR Battle Royale]{Two in-game screenshots of the game Stand out: VR Battle Royale show the two video perspectives used in the survey. The upper screenshot shows the game world from the first-person perspective, depicting a room with a table and an open door that offers a view of a wide outdoor area. The screenshot below shows the same game scene, but additionally includes a recording of the real player who is blended into the game environment.}
\end{figure}

\textit{Stand Out} is a first-person shooter in which the player plays online against a large group of other players  (see Figure~\ref{fig:standout}). Following the battle royale principle, the goal is to be the last survivor on the island where the game takes place. To win, the player has to collect weapons and ammunition, shoot other players, and move across the island. The player can travel larger distances using the control stick. Hence, in contrast to the other two games, locomotion is very prevalent in \textit{Stand Out}. Like in other first-person shooters, the gameplay is rather fast-paced in general. The focus of attention is very dynamic since the player has to react quickly in case of an attack.

All in all, the three games differ mainly concerning pace, focus, and locomotion. We assume that these characteristics can potentially influence the spectators' experience in the two perspectives under investigation, as these provide different main focal points. For instance, quick game events might be more comprehensible for spectators if they see both the player and the game world in the third-person mixed-reality perspective. On the other hand, the first-person perspective might be more suitable for games with a dynamic focus and much locomotion. Therefore, we included all three games in both perspectives in our study to investigate potential differences.

\subsection{Implementation of the Different Perspectives}
\label{sec:implementation}
Overall, there are several possibilities to compile a stream of a VR gaming session. We decided to compare two basic approaches which are commonly used by streamers and at the same time differ significantly regarding their main focal point: the first-person view and a mixed-reality third-person perspective. While some streamers also use a combination of different views by compiling picture-in-picture modes, we focus on the two main approaches, as we are particularly interested in how spectators evaluate the possibility to see the player integrated in the game world in the third-person view and the lack thereof in the first-person view.

As stated above, we recorded two gameplay videos for each of the three games: one with the first-person perspective and one with the third-person perspective. In all cases, the same player (male, 26 years old) played the game and all videos are about three minutes long. 

The first-person view was simply a screen-recording of the game from the player's point of view.
To create the third-person mixed-reality views, we used the software \textit{LIV}~\cite{LIV}, which allows integrating a green screen recording of the player into the game world. We used an additional, static game camera to capture the game scene from behind the player (cf. Figures~\ref{fig:beatsaber}, \ref{fig:superhot}, and \ref{fig:standout}). We also considered rotating the mixed-reality camera dynamically based on the player’s actions. While this approach is technically possible using \textit{LIV}, it requires a far more sophisticated setup. Since a dynamic camera was not required to address our main research question and since we wanted to stick to the most commonly used techniques in the gaming community, we discarded this option.
Instead, we tested different positions while implementing the third-person views to find an appropriate static camera position for each game.

\textit{LIV} also enables to replace the real player by an avatar model, a feature used by some streamers, as well. However, such a third-party avatar is not visually matched to the game and, thus, introduces an additional source of interference. While the real player's appearance also mismatches with the game world, a mixed-reality view best reveals the manipulations conducted by the player in the direct context of the game. For these reasons, we decided to not use a virtual avatar.

\subsection{Study Plan and Survey Structure}
We conducted a mixed design online study with the game shown in the videos as a between-subjects variable and the video perspective as a within-subjects variable. 
That means, each participant was randomly assigned to one game and watched both videos of that game. The order of the two perspectives was counterbalanced, as well, to avoid bias due to potential sequence effects.

The survey started with a short introduction, informing participants about the goal, procedure, and anonymity of the study. Then we asked for basic demographic data, including age, gender, and nationality. Additionally, we requested some information about participants' familiarity with VR headsets and VR games, as well as their digital gaming and streaming habits.

As we were also interested in viewers' general motivations to watch videos or streams of VR games, we compiled a list of possible motives based on the uses and gratification theory. More precisely, we derived our items from the work of Sjöblom et al.~\cite{sjoblom2017people}, who investigated the motivation of Twitch users. 
Although this motivational model does not explicitly refer to VR streaming content, we believe that the general types of motives of viewers are largely independent from the platform used by the streamer (including VR setups). Hence, we think the model includes all high-level motivations that are relevant in the context of our study.
The question and the final list of answers can be found in Table~\ref{tab:UGOverview}. Participants were asked to select all reasons that apply (multiple answers were possible). We also included the option \textit{"None (I would not watch a video of a VR game)"}, to be able to identify participants having no interest in the study's topic.

\begin{table*}[ht]
    \caption{Overview of the different motivational aspects related to the viewing of VR game videos (participants were asked to select all answers that apply).} 
  \begin{tabularx}{\textwidth} {>{\raggedright\arraybackslash}p{3.2cm}>{\raggedright\arraybackslash}X >{\centering\arraybackslash}p{1cm}}
    \toprule
    \addlinespace
    \textbf{Class of Gratification (based on Sjöblom et al.~\cite{sjoblom2017people})}  & \textbf{Which of the following reasons could motivate you to watch a video where a player is playing a VR game?} & \textbf{Votes} (\textit{N}=217)  \\ 
    \midrule
\addlinespace
     cognitive & to inform me about the game or to get an impression of it (\textit{information seeking})       
     &  113    \\
     cognitive & to learn new game strategies or how to master the game   (\textit{learning game strategies})    
     & \ 84   \\
     affective & because it is entertaining and/or exciting (\textit{enjoyment})
     &  119   \\
    % because it is fun 
    % &  106   &    \\
      personal integrative & to be able to comment and have a say (\textit{recognition})
     & \ 24   \\
     social integrative & in order not to feel alone (\textit{companionship})
     & \ 26    \\
     tension release & to distract me and pass the time (\textit{distraction})
     & \ 62  \\
     tension release & in order to relax (\textit{relaxation})
     & \ 46   \\
  \bottomrule
\end{tabularx}
\label{tab:UGOverview}
\Description[Cognitive and affective gratifications are voted most often]{Table 1 gives an overview of the different motivational aspects related to the viewing of VR game videos. The most prominent gratifications relate to the categories cognitive (information seeking and learning game strategies) and affective (enjoyment).}
\end{table*}

Following this first, general part of the questionnaire, we asked participants to ensure that their speakers or headphones are active to be able to hear the sound of the videos and then showed them the first video. To control that the video was not forwarded or skipped, we measured the time participants spent with the video. This way, we were able to identify participants who skipped (parts of) the videos and label their data as invalid.

After the video, we administered the enjoyment subscale of the Intrinsic Motivation Inventory (IMI)~\cite{ryan2000self} to assess how much participants enjoyed watching the video.
To further investigate the viewing experience, we asked additional custom questions about the view and the comprehensibility of the video, as well as the perceived involvement. The full list of questions can be found in Table~\ref{tab:ViewExperience}.
Then the second video was shown, and again IMI and the custom questions were administered after that. Then, we asked whether participants knew or have played the game shown in the videos before, how much they like the game and how much they like the genre it belongs to in general.
Finally, participants were asked which of the two perspectives they preferred. There was also the option to indicate that they did not have a preference. In a free-text form, we asked participants to give reasons for their decision. Moreover, participants could provide any additional notes.

Considering that we cannot completely control the setting and conditions under which participants take part in an online study, we increased the validity of the data by including sanity check questions. For this purpose, we asked the same question twice with reversed scales, to ensure that participants have read the question text and did not select random answers.

\subsection{Recruitment and Sample}
We were interested in the opinion of potential spectators of VR game videos and aimed at improving their viewing experience. Thus, we defined all persons who have at least some interest in VR technology and digital games as our target group, with no further restrictions regarding demographic data or prior experience with VR. We promoted the survey on different online channels, both in English and in German. That included several Reddit communities and Facebook groups related to the topics game streaming or VR games. Moreover, we also used more general groups that are aimed at the recruitment of online survey participants.

In total, 316 participants completed the survey. Sixty-nine of these cases had to be excluded from the analysis because participants failed the sanity check questions or did not watch the videos completely. Moreover, we excluded 30 additional participants, who stated that they had no interest in VR games and would never view videos of such games voluntarily. Those participants do not match our target group. Hence, our final sample contains 217 participants.

The sample includes a wide variety of nationalities (27 different countries), with 63 German and 77 American participants being the majority. 
The mean age of participants was 28 (\textit{SD}~=~8.69), with a range from 16 to 64. Regarding gender, the sample included 125 male and 92 female participants. About three-quarters of all participants (\textit{N}~=~172) reported that they regularly played digital games.  Many participants also had prior experience with VR games, with only 58 persons stating that they have not yet used a VR headset. Regarding the question how often they watched gaming videos/streams on average, most participants (\textit{N}~=~189) reported that they watched game streams at least once a month.

Concerning our three game subgroups, the distribution is a bit uneven: 89 participants viewed the videos of \textit{Superhot VR}, 67 participants viewed \textit{Beat Saber}, and 61 \textit{Stand Out}.
However, the distribution of age, gender, and nationality is comparable among the three groups.
About two thirds of the participants in the \textit{Beat Saber} group knew the game before (\textit{N}~=~44) and nearly half of the group had played the game themselves (\textit{N}~=~29). \textit{Superhot VR} was known to half of the participants (\textit{N}~=~40) and 33 participants had played the game. \textit{Stand Out} was less known by our participants, with only 13 persons being familiar with the game, of whom 8 had played it. Asked about how much they liked the game, participants in all three game groups rated the games slightly positive on average on a scale from 0 to 6 (Beat Saber: \textit{M}~=~4.19, \textit{SD}~=~1.79; \textit{Superhot VR}: \textit{M}~=~3.53, \textit{SD}~=~2.07; \textit{Stand Out}: \textit{M}~=~3.15, \textit{SD}~=~1.99). Similarly, participants stated to rather like the genre of the game they watched in general (Beat Saber: \textit{M}~=~4.39, \textit{SD}~=~1.65; \textit{Superhot VR}: \textit{M}~=~3.73, \textit{SD}~=~2.17; \textit{Stand Out}: \textit{M}~=~3.30, \textit{SD}~=~2.13).

\section{Results}
In the first step of our data analysis, we have a look at participants' general motivation to watch videos of VR games.
After that, we address our main research question by comparing participants' evaluations of both video perspectives and their preferences.

\subsection{Motivation to Watch VR Game Videos}

To examine participants' general motivation to watch VR game videos, we analyzed their answers to the uses and gratifications question. Table~\ref{tab:UGOverview} shows how often participants selected each reason to watch a VR game video.  
Whereas all different motivations received some votes, the distribution of votes indicates that affective and cognitive gratifications were most prevalent among our participants. The majority of participants would watch videos of VR games, if they seek information about the game (\textit{N}~=~113) or because they enjoy watching them and feel entertained (\textit{N}~=~119).
Learning about game strategies was also mentioned often (\textit{N}~=~84). On the other hand, fewer participants see tension release, both in terms of distraction (\textit{N}~=~62) and relaxation (\textit{N}~=~46), as a motivation to watch videos of VR games. Finally, personal and social integrative motives were least prevalent (\textit{N}~=~24 and \textit{N}~=~26).

\begin{figure*}
\centering
\includegraphics[width=\textwidth]{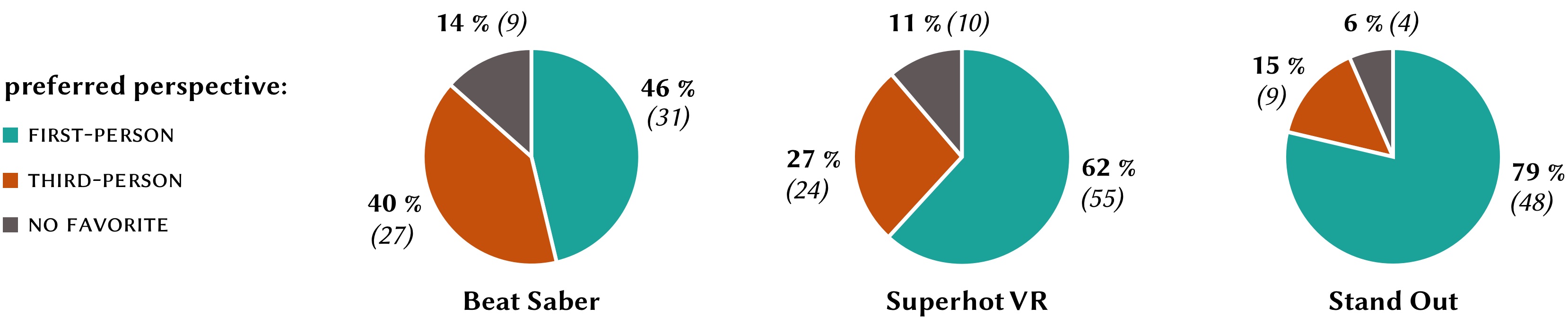}
\caption{Distribution of the preferred perspective votes of our participants for the three games \textit{Beat Saber}, \textit{Superhot VR}, and \textit{Stand Out}. }
\label{fig:piechart}
\Description[Superhot VR and Stand Out show clear preferences for the first-person perspective, whereas for Beat Saber votes are more evenly distributed]{Three pie charts show which perspective was preferred by the participants of our study in the three different games Beat Saber, Superhot VR and Stand Out. There were three possible answers, namely 'first-person', 'third-person', and 'no favorite'. In the Beat Saber condition, the first-person and the third-person perspective received comparable number of votes. In contrast, Superhot VR and Stand Out show clear preferences for the first-person perspective.}
\end{figure*}

\subsection{Evaluation of First- and Third-Person Perspectives}

With regard to our main research question, we analyze how participants perceived both perspectives and which one they preferred.  

Overall, the voting shows a recognizable preference for the first-person perspective: 134 of all 217 participants preferred the first-person perspective, whereas only 60 voted for the third-person perspective. Twenty-three participants stated that they have no favorite view. 
We performed Pearson chi-square tests to investigate if there are significant relations between the preferred perspective and certain characteristics of participants that might influence their vote, namely their gender, whether or not they were familiar with the game that was shown, and their general motivations to watch VR game videos. 
Regarding gender, vote distribution is very similar between male and female participants, and there is no significant correlation, ${\chi}^2$(2)~=~2.31, \textit{p}~=~.316.
Participants' familiarity with the game seems to have no effect on their voting, either, ${\chi}^2$(2)~=~0.35, \textit{p}~=~.839.
To test the influence of different general motivational aspects, we performed chi-square tests for each of the statements shown in Table~\ref{tab:UGOverview}. The results indicate that whether or not participants selected a particular motivation is not related to their preferred perspective, as no significant correlations could be found (all \textit{p}~>~.348).

As there might be differences with regard to the three games we tested, we further investigate participants' preferences in the three subgroups for \textit{Beat Saber}, \textit{Superhot VR}, and \textit{Stand Out}. Therefore, we split our data for the following analysis and report results for each game individually. 
Figure~\ref{fig:piechart} shows participants' preferred perspective in the three study conditions. In line with the overall result, the first-person perspective received most votes in all cases. However, there is a noticeable difference regarding the distribution of votes: whereas there is a clear preference for the first-person perspective in the \textit{Stand Out} group (48 out of 61 participants) and the \textit{Superhot VR} group (55 out of 89 participants), the votes in the \textit{Beat Saber} group are almost evenly distributed with 31 participants preferring the first-person perspective and 27 participants preferring the third-person perspective.
A chi-square test underlines that the game that was shown had a significant influence on the vote of the preferred perspective, ${\chi}^2$(4)~=~14.48, \textit{p}~=~.006, Cramer's \textit{V}~=~0.183 (no expected cell frequencies were below 5). Though the effect size is only small (\textit{V}~<~0.3), the result indicates that the two video perspectives were perceived differently in \textit{Beat Saber} than in the other games.

To investigate the reasons why participants prefer one view to the other, we compared the viewing experiences between both perspectives and tested for significant differences. Table~\ref{tab:ViewExperience} shows all mean values for both perspectives in the three game conditions. For each game and each dimension of the viewing experience, we performed repeated measures analysis of variance (ANOVA) with perspective as a within-subjects variable and order of game views as a between-subjects factor to test for potential sequence effects. In the following, we report the results of these analyzes for each game. In the interest of better legibility, we only report on sequence effects if they are significant. If not mentioned, the analysis did not show a significant interaction effect between the experience dimension and the order of the two perspectives.

\begin{table*} 
    \caption{Mean values and standard deviations (\textit{M}(\textit{SD})) of different aspects of the viewing experience in the three study conditions (games) \textit{Beat Saber}, \textit{Superhot VR}, and \textit{Stand Out}, comparing the first-person and the third-person perspectives. Each item was rated on a 7-point scale ranging from 0 to 6. Significant differences between two perspectives are indicated in bold print.} 
  \begin{tabularx}{\textwidth} 
  {>{\raggedright\arraybackslash}p{4.5cm} 
  >{\raggedleft\arraybackslash}X
  >{\raggedright\arraybackslash}X
  >{\raggedleft\arraybackslash}X
  >{\raggedright\arraybackslash}X
  >{\raggedleft\arraybackslash}X
  >{\raggedright\arraybackslash}X}
    \toprule 
    \addlinespace
   &  \multicolumn{2}{c}{\textbf{Beat Saber} (\textit{N}=67)} 
   & \multicolumn{2}{c}{\textbf{Superhot VR} (\textit{N}=89)}  
   & \multicolumn{2}{c}{\textbf{Stand Out} (\textit{N}=61)}  \\

    & 1st person & 3rd person 
    & 1st person & 3rd person  
    & 1st person & 3rd person \\
    \midrule
\addlinespace
    \textbf{IMI} \\
     \ \ \ \ \ \ Enjoyment        
     &  3.18 (1.47)  &  3.26 (1.44) 
     &  3.03 (1.70) & 3.10 (1.64)
     &  \textbf{3.21} (1.52) & \textbf{2.43} (1.73)\\
     \addlinespace
     \textbf{Focus and Clear View} \\
     F1) I saw well how the player  
         %Insight2    
     &  4.24 (1.61) & 4.57 (1.46) 
     &  4.28 (1.60) & 4.10 (1.62)
     &  \textbf{4.54} (1.21) & \textbf{3.74} (1.70)\\
     \ \ \ \ \ \ interacted with objects in the \\
      \ \ \ \ \ \  game world. \\
     F2) While watching, I had the           
     &  2.58 (1.83)  &  2.19 (1.78) 
     &  \textbf{2.31} (1.97)  &  \textbf{3.03} (2.07)
     & \textbf{2.70} (1.80) & \textbf{3.93} (1.89)\\
     \ \ \ \ \ \ feeling of missing important \\
     \ \ \ \ \ \ things in the game because I  \\
     \ \ \ \ \ \ couldn't see them. \\
     F3) I had a good view of the game         
     &  4.15 (1.49)  &  4.24 (1.46) 
     &  \textbf{4.18} (1.47)  &  \textbf{3.65 }(1.72)
     &  \textbf{4.14} (1.38) & \textbf{3.13} (1.94)\\
      \ \ \ \ \ \ world. \\
     \addlinespace
     \textbf{Comprehensibility} \\
     C1) I always understood what       
     &   4.52 (1.58)  &  4.75 (1.47) 
     &   \textbf{4.33} (1.43)  & \textbf{3.80} (1.63)
     &   \textbf{4.23} (1.33)  &  \textbf{3.64} (1.89)\\
     \ \ \ \ \ \ happened in the game. \\
     C2) At any time I could          
     &   3.93 (1.64)  &   4.42 (1.63) 
     &   4.17 (1.51)  &  4.06 (1.74)
     &   \textbf{4.44} (1.35)  &  \textbf{3.78} (1.75) \\
      \ \ \ \ \ \ comprehend what the player  \\
       \ \ \ \ \ \ was doing in the VR world. \\
     C3) I was able to understand how          
     &  4.33 (1.66)  &   4.61 (1.59) 
     &  \textbf{4.37} (1.58)  &   \textbf{3.85} (1.70)
     &  \textbf{4.10} (1.47)   &  \textbf{3.16} (1.91) \\
     \ \ \ \ \ \ successful the player was in  \\
     \ \ \ \ \ \ the game. \\
     \addlinespace
     \textbf{Involvement} \\
     I1) \ I felt like being part of the 
     &  \textbf{3.06} (1.88)  &   \textbf{2.46} (1.97) 
     &  \textbf{3.28} (1.85)  &   \textbf{2.35} (1.95)
     &  \textbf{3.67} (1.88)  &   \textbf{2.54} (2.28)\\
      \ \ \ \ \ \ game. \\
     I2) \ I saw the virtual world as if I         
     &  \textbf{3.54} (1.99)  &  \textbf{2.70} (1.92) 
     &  \textbf{3.35 }(1.85)  &  \textbf{2.44} (1.89)
     &  \textbf{3.93} (1.83)  &  \textbf{2.59} (2.03)\\
     \ \ \ \ \ \ was there myself. \\
  \bottomrule
\end{tabularx}
\label{tab:ViewExperience}
\Description[While there are few significant differences in the viewing experience for the game Beat Saber, every aspect differs significantly between the two perspectives in the game Stand Out]{This table list all mean values and standard deviations of the different aspects of the viewing experience in the three study conditions (games) Beat Saber, Superhot VR, and Stand Out comparing the first-person and the third-person perspectives. Significant differences are highlighted in bold print. While there are few significant differences for the game Beat Saber, every aspect differs significantly between the two perspectives in the game Stand Out.}
\end{table*}

\subsubsection{Beat Saber}
In the \textit{Beat Saber} group, most differences are not significant. Neither enjoyment nor any ratings of focus and clear view and comprehensibility were rated significantly different between the two perspectives (all \textit{p}~>~.05). In contrast, the two questions regarding the perceived involvement of the viewers show significant differences: in the first-person view, participants felt more like being part of the game (I1), \textit{F}(1,~65)~=~6.06, \textit{p}~=~.016, and like being in the virtual world (I2), \textit{F}(1,~65)~=~9.04, \textit{p}~=~.004.

\subsubsection{Superhot VR}
In the \textit{Superhot VR} condition, the repeated measures ANOVA revealed more significant differences. In the first-person perspective, the ratings regarding having a good view of the game world (F3) were higher than in the third-person perspective, \textit{F}(1,~87)~=~5.64, \textit{p}~=~.020. Additionally, the feeling of missing important things (F2) was significantly higher in the third-person perspective, \textit{F}(1,~87)~=~6.93, \textit{p}~=~.010.
In terms of comprehensibility, participants had the feeling of significantly better understanding what happened in the game (C1), \textit{F}(1,~87)~=~7.47, \textit{p}~=~.008, and how successful the player was (C3), \textit{F}(1,~87)~=~7.93, \textit{p}~=~.006, in the first-person perspective.
Similar to the results in the \textit{Beat Saber} group, both items regarding involvement (I1 and I2) were rated significantly higher in the first-person perspective,  
\textit{F}(1,~87)~=~17.19, \textit{p}~<~.001 (I1), and \textit{F}(1,~87)~=~15.91, \textit{p}~<~.001 (I2).
However, in the \textit{Superhot VR} condition, there was also a significant interaction effect between the ratings for involvement and the order in which the two videos were watched, indicating sequence effects. The rating for being part of the game (I1) was particularly high for the first-person perspective if participants had viewed the third-person perspective video beforehand (\textit{M}~=~3.64 compared to \textit{M}~=~2.88),
\textit{F}(1,~87)~=~6.27, \textit{p}~=~.014.
The same pattern becomes apparent for the item \textit{"I saw the virtual world as if I was there myself"} (\textit{M}~=~3.98 compared to \textit{M}~=~2.64), \textit{F}(1,~87)~=~13.40, \textit{p}~<~.001.
All other differences (IMI, F1, and C2) were not significant (all \textit{p}~>~.05).

\subsubsection{Stand Out}
In the \textit{Stand Out} group, enjoyment (IMI) was significantly higher in the first-person video, \textit{F}(1,~59)~=~13.68, \textit{p}~<~.001. 
Moreover, participants gave significantly better ratings regarding focus and clear view in the first-person perspective: they better saw how the player interacted with game objects (F1), \textit{F}(1,~59)~=~12.47, \textit{p}~=~.001, and had a better view of the game world (F3), \textit{F}(1,~59)~=~15.53, \textit{p}~<~.001. At the same time, the feeling of missing important things was lower (F2), \textit{F}(1,~59)~=~14.43, \textit{p}~<~.001.
The three items regarding comprehensibility were also rated significantly higher in the first-person perspective: participants better understood what happened in the game (C1), \textit{F}(1,~59)~=~4.10, \textit{p}~=~.047, what the player was doing (C2), \textit{F}(1,~59)~=~6.98, \textit{p}~=~.011, and how successful the player was (C3), \textit{F}(1,~59)~=~16.60, \textit{p}~<~.001. 
In line with the other two groups, involvement (I1 and I2) was significantly higher in the first-person perspective: \textit{F}(1,~59)~=~16.39, \textit{p}~<~.001 (I1), and \textit{F}(1,~59)~=~26.23, \textit{p}~<~.001 (I2).
Summarized, all aspects of the viewing experience differ significantly between the first- and the third-person perspective in the \textit{Stand Out} group, with the first-person perspective being rated better in all cases.

\begin{table*}
    \caption{Results of the thematic analysis with regard to reasons why participants preferred the first-person perspective. The middle column contains exemplary quotes of participants which were assigned to the topics. The right column shows the number of mentions, i.e. how many single answers of participants were assigned to the respective topic.} 
  \begin{tabularx}{\textwidth} {>{\raggedright\arraybackslash}p{0.1cm}>{\raggedright\arraybackslash}X >{\raggedright\arraybackslash}p{6.7cm}>{\centering\arraybackslash}p{1.4cm}}
    \toprule
    \addlinespace
    \multicolumn{2}{l}{\textbf{Reasons to Prefer the }}  & \textbf{Examples} & \textbf{Mentions}\\
    \multicolumn{2}{l}{\textbf{First-Person Perspective}}  &  \\
    \midrule
\addlinespace
     \multicolumn{2}{l}{\textbf{Involvement}} & 
     \\   
     &  The viewers felt more immersed, they felt like being part of the game, being in the game world, or being the player.
     & \textit{I like the first-person perspective because it makes me feel like I'm playing the game, not someone else. It is more entertaining when I feel like I'm part of the game.}
     & 38\\
     \addlinespace
      \multicolumn{2}{l}{\textbf{Focus}} 
     \\   
     &  The viewers think that the focus was better, because they were able to see all important things and did not miss something outside the viewport.
     & \textit{It gives me the ability to see the important parts of the game as they happen, rather than being stuck facing one direction, missing details that are behind my point of view.}
     & 12\\
      \addlinespace
      \multicolumn{2}{l}{\textbf{Comprehensibility}} 
       \\
     & The viewers better understood what happened in the game and what the player was doing.
     & \textit{First person (in this game at least) lets viewers understand what the player is doing.}
     & 10\\
      \addlinespace
      \multicolumn{2}{l}{\textbf{Obstructive Player in Third Person}} 
         \\
     & The player in the third-person view was perceived as obstructive, because he obscured the view on the game world and did not fit to the environment. 
     & \textit{First person allows for better visibility without obstruction by the player.}
     & 10\\
      \addlinespace
      \multicolumn{2}{l}{\textbf{Original Game Perspective}}  
      \\
     & The first-person view corresponds with the original game perspective, hence viewers can better imagine how it would be to play the game. 
     & \textit{I don't like the mixed reality view. I want to see exactly what the player sees.}
     &8 \\
      \addlinespace
      \multicolumn{2}{l}{\textbf{Realism}} 
        \\
     & The experience felt more real to viewers.
     & \textit{Because it looks more real to me.} 
     & 7 \\
  \bottomrule
\end{tabularx}
\label{tab:ThematicAnalysis1st}
\Description[Six identified reasons to prefer the first-person perspective: involvement, focus, comprehensibility, obstructive player, original game perspective, and realism]{The table lists the results of the thematic analysis with regard to reasons why participants preferred the first-person perspective. The first column shows the six identified topics involvement, focus, comprehensibility, obstructive player, original game perspective, and realism. The middle column contains exemplary quotes of participants which were assigned to the topics. The right column shows the number of mentions, i.e., how many single answers of participants were assigned to the respective topic.}
\end{table*}

\begin{table*}[ht]
    \caption{Results of the thematic analysis with regard to reasons why participants preferred the third-person perspective. The middle column contains exemplary quotes of participants which were assigned to the topics. The right column shows the number of mentions, i.e. how many single answers of participants were assigned to the respective topic.} 
  \begin{tabularx}{\textwidth} {>{\raggedright\arraybackslash}p{0.1cm}>{\raggedright\arraybackslash}X >{\raggedright\arraybackslash}p{6.7cm}>{\centering\arraybackslash}p{1.4cm}}
    \toprule
    \addlinespace
    \multicolumn{2}{l}{\textbf{Reasons to Prefer the }}  & \textbf{Examples}  & \textbf{Mentions} \\
    \multicolumn{2}{l}{\textbf{Third-Person Perspective}}  &  \\
    \midrule
\addlinespace
     \multicolumn{2}{l}{\textbf{Player's Movements}} \\
     & To see the player's movement and his interaction with the game world is more entertaining and interesting. 
     & \textit{It was more interesting to see how the person was actually moving around and how it looked like he was actually in the game world.} & 19\\
     \addlinespace
      \multicolumn{2}{l}{\textbf{Comprehensibility}} \\
     & The viewers better understood what the player was doing.
     & \textit{It was easier to see what the player was doing in the game world.} & 11\\
      \addlinespace
      \multicolumn{2}{l}{\textbf{Motion Sickness in First Person}}  \\
     & It was more comfortable, because in the first-person view viewers experienced dizziness or nausea.
     & \textit{Watching first person made me kind of dizzy so the third-person perspective was more interesting and more comfortable to watch.} & 6\\
      \addlinespace
      \multicolumn{2}{l}{\textbf{View on Game World}} \\
     &  The viewers feel that they can see more of the game world.
     & \textit{The third-person perspective gave me a wider view of the world in which the game was taking place.} & 4\\
    
  \bottomrule
\end{tabularx}
\label{tab:ThematicAnalysis3rd}
\Description[Four identified reasons to prefer the third-person perspective: player's movements, comprehensibility, motion sickness, and view on game world]{The table lists the results of the thematic analysis with regard to reasons why participants preferred the third-person perspective. The first column shows the four identified topics player's movements, comprehensibility, motion sickness, and view on game world. The middle column contains exemplary quotes of participants which were assigned to the topics. The right column shows the number of mentions, i.e., how many single answers of participants were assigned to the respective topic.}
\end{table*}

\subsection{Thematic Analysis: Reasons for Preferred Perspective} 

To gain further insight into the positive and negative qualities of the two perspectives, we performed a thematic analysis of the free text answers to the question of why participants prefer one perspective to the other. 
For this purpose, two researchers looked at all answers independently and sorted them by recurring topics. We followed a deductive approach based on the reflexive thematic analysis described by Braun and Clarke~\cite{braun2006}, with an additional check of inter-rater agreement. After the first round of clustering, both researchers compared their lists and discussed all differences. Based on the discussion, a final clustering was agreed upon.

We identified six clusters that describe reasons why participants preferred the first-person perspective, as shown in Table~\ref{tab:ThematicAnalysis1st}. 
Many participants (\textit{N}~=~38) highlighted a higher involvement perceived in the first-person perspective. They reported that this perspective made them feel like being part of the game or even being the player themselves.
Besides, some participants (\textit{N}~=~12) pointed out that the focus was better in the first-person perspective because they were able to see the important things (such as enemies approaching). Participants also reported that the comprehensibility was higher, as they were better able to follow the game events (\textit{N}~=~10). In the third-person perspective, the player was perceived as an obstacle by some participants (\textit{N}~=~10), covering parts of the game and interfering with immersion.
Apart from the higher involvement, some participants (\textit{N}~=~8) also emphasized that they prefer the first-person perspective because it is the \textit{"original game perspective"}. This way, they experience how the game looks to the player and can better imagine how it would feel to play the game. Finally, some participants (\textit{N}~=~7) pointed out that the experienced realism was higher in the first-person perspective.

For the third-person perspective, we identified four categories of reasons to prefer it to the first-person perspective, as shown in Table~\ref{tab:ThematicAnalysis3rd}.
The most frequently mentioned reason was that participants (\textit{N}~=~19) liked to see the player and his movements. They reported that it was more interesting and entertaining to focus the player and to be able to observe the direct interaction between the player and the game's environment. 
Related to seeing the player's movement, some participants (\textit{N}~=~10) also highlighted that they gained a better understanding of how the game is played and how the interaction works. Hence, they stated that the comprehensibility was better in the third-person perspective. 
Besides, some participants (\textit{N}~=~6) preferred the third-person view, because they experienced some form of motion sickness in the first-person perspective. They reported that it was more comfortable to watch the game in third person. Finally, some participants (\textit{N}~=~4) preferred the third-person perspective, because they think that it enabled them to see more of the game world.

Part of the identified reasons to prefer one perspective to the other were mentioned comparably often in all three game groups. More precisely, participants in each group addressed the topics higher involvement, realism, and the original game perspective of the first-person perspective, as well as less motion sickness and a better view on the game world in the third-person perspective. In contrast, some topics were more prevalent for specific games. The better focus and the better comprehensibility of the first-person perspective were predominantly mentioned by participants who had watched the videos of \textit{Superhot VR}: we counted focus eight times and comprehensibility six times in the \textit{Superhot VR} condition, while both topics appeared only two times in each of the other two conditions. However, at this point we want to remind that the \textit{Superhot VR} group was also bigger than the other two groups (\textit{N}~=~89 vs. 67 and 61), which might account for such differences.

In the \textit{Stand Out} condition, participants complained more about the obstructive player in the third-person perspective (N~=~8) than participants in the other two groups. 
Moreover, two reasons to prefer the third-person perspective---seeing the player's movements and comprehensibility---were mentioned for both \textit{Beat Saber} and \textit{Superhot VR}, but not for the game \textit{Stand Out}. Even though the Stand Out group was a bit smaller than the other two study groups, the difference is still noticeable.

\section{Discussion}

We observed an overall preference for the first-person perspective in our study. However, we found significant differences between the three games. This result confirms our assumption that the choice of an appropriate perspective is dependent on the particular game. Moreover, the perceived benefits and shortcomings of both perspectives as reported by our participants indicate that personal preferences and the motivation of the viewer also play an important role. 

\subsection{Influence of Game Characteristics on Perspective Preferences} 
We received the most homogeneous feedback for the game \textit{Stand Out}. Very few participants preferred the third-person perspective, and it also performed significantly worse regarding all measured aspects of the viewing experience, including overall enjoyment. Many participants mentioned a feeling of confusion and the impression of missing essential parts of the gameplay. Whereas viewers of the other two games rather appreciated seeing the player in action according to our thematic analysis, viewers of \textit{Stand Out} experienced the player as obstructive in the third-person view.
We assume that this issue is caused by a mismatch between the focus of the viewer in the third-person perspective, which lies on the player, and the location of the important game events: in \textit{Stand Out}, the main actions---such as approaching enemies, the search for coverage or gun fights---are not centered around the player's position, but evolve further away in the surrounding.
A first-person perspective better matches this game characteristic and, thus, seems to be more appropriate for this kind of games.

For the game \textit{Superhot VR}, the participants were able to perceive the player's actions and interactions with objects in both views equally and preferences are less clearly distributed.
In contrast to \textit{Stand Out}, there is no significant difference in the IMI enjoyment subscale: both perspectives induced similar levels of enjoyment. Since entertainment was the most commonly mentioned motivator to watch VR game videos, we can assume that at least some participants preferred the third-person view in \textit{Superhot VR} for enjoyment reasons.
Nevertheless, many participants still disliked the third-person perspective due to the feeling of having a limited view and missing important game events.
Comments of some participants point towards a possible explanation: these viewers explicitly stated problems with situations where the player reacted to opponents that were not visible on the screen. This issue seems similar to the problems reported for the third-person perspective in \textit{Stand Out}. Yet, the problem is less prominent in \textit{Superhot VR} and only applies to certain situations. In contrast to \textit{Stand Out}, \textit{Superhot VR} also contains important game events that are directly centered around the player, such as dodging attacks in slow-motion. Such events might account for the fact that still 27\% of participants preferred the third-person perspective, which offers a good view on the player. We assume that a more dynamic third-person camera could reduce the issue of the limited view and increase approval of the third-person view to a certain extent.

The most inconclusive results are the ones for the game \textit{Beat Saber}. In this case, our sample shows no clear preference for one perspective. 
Regarding the spectator experience, only involvement---measuring the feeling of being part of the game---was rated higher for the first-person view. All other subscales, namely enjoyment, comprehensibility, and seeing everything that is important, do not indicate any difference between the two perspectives.
These results indicate that the third-person view seems to have advantages to this particular game and that the first-person alternative is not preferable in every case. \textit{Beat Saber} seems to be more appropriate for the third-person perspective than both other games.  

Considering the identical study conditions and similar audiences, the reasons for the measured differences between all three games have to reside within the particular game characteristics. Our results indicate that the focus on the player's bodily interaction in the third-person perspective is more compelling for spectators of \textit{Beat Saber} than for viewers of the other two games. 
In contrast to \textit{Stand Out} and \textit{Superhot VR}, \textit{Beat Saber} requires very fast and coordinated movements of the players. All relevant game events (approaching blocks and hits of the player) are tightly coupled to these movements both temporally and visually. Watching this type of experience is likely more interesting if the viewers can see the players and their movements, as they have an immediate effect on the gameplay. Spectators of the other two games might prefer the first-person perspective, because the player's bodily interaction looks less intriguing and, hence, a clear focus on the in-game events is more interesting.

In addition, the overall high pace of the players' movements in \textit{Beat Saber} makes it hard for new viewers to understand and follow the gameplay. In this case, the third-person perspective could help the audience to gain a better understanding of the game and its goals. For the other two games, it is more important to see the players' view and their interactions with the weaponry to understand the overall gameplay and the players' strategies. 

Moreover, we assume that the fixed viewing direction of \textit{Beat Saber} contributes to the success of the third-person view. During the game, the player's view is mostly fixed in one direction, which makes it easy to align the third-person camera with the main course of action. As a result, the viewers' impression of missing essential aspects is reduced. For comparison, \textit{Superhot VR} features a more dynamic environment where enemies approach the player from multiple directions. \textit{Stand Out} provides the most dynamic locomotion system that combines virtual motion and rotation with real movements. Additionally, it relies heavily on long travel distances.

In summary, we assume that the key difference between the three games is the visual coupling of the main game actions and the player's position and movement. In games like \textit{Beat Saber} all relevant game events are centered directly around the player and, thus, emphasized by the third-person-perspective. In games like \textit{Stand Out} most events dynamically evolve in the wider surroundings. In the latter case, the first-person perspective is more appropriate, as it better guides the focus of the spectator towards the important game events.

Despite the discussed reasons that explain the usefulness of the third-person perspective for games like \textit{Beat Saber}, a considerable number of our participants still favored the first-person view for this game. This preference hints towards certain desires of the spectators---such as experiencing the game from the player's view---that require a first-person perspective and are less linked to characteristics of the game. This finding is especially interesting considering that using a mixed-reality third-person view is a widespread approach in current \textit{Beat Saber} videos and streams. 

\subsection{Subliming the Perceived Strengths and Weaknesses of Both Perspectives}
Our analysis of the three VR games has shown that certain game characteristics seem to influence the suitability of the two different streaming perspectives. However, we also consider spectators' personal preferences and motives to be a relevant factor. 
The UG results revealed two primary motivators of our participants for watching VR game videos: entertainment and information seeking. This finding fits the most frequently mentioned reasons for choosing one perspective over the other: involvement and comprehensibility. 

Whereas we could not identify significant correlations between participants' general motives and the perspective they preferred, our thematic analysis of participants' reasons to prefer a certain perspective further helps to understand the perceived strengths and weaknesses of both views. The first-person view is preferred by spectators who want to feel like they are playing the game themselves and who want to see \textit{"through the player's eyes"}. Some participants explicitly mentioned a \textit{"preference for the original perspective"}. An increased involvement was the most prevalent reason of our participants to prefer the first-person perspective. This finding of the thematic analysis is underlined by our questions concerning the feeling of \textit{"being part of the game"}. For all three games, participants felt significantly more as a part of the game in the first-person view. Hence, the first-person perspective better fosters immersive experiences of spectators than the third-person view.

On the downside, some participants indicated that \textit{"seeing through the player's eyes"} made them dizzy and motion sick. In these cases, participants preferred the third-person view, which seems to be less prone to motion sickness.

Interestingly, a better \textit{"comprehensibility"} is mentioned as an perceived advantage for both views. Participants disagreed which perspective provides a better understanding. This feedback might be the result of the different foci of both perspectives: In the first-person perspective, viewers experience the game exactly how it would look like playing it. Hence, some participants might have the feeling that this view provides a better overall impression of the game. In the third-person view, the spectators see the player's movements and the resulting actions in the virtual environment. They might feel that this matching between manipulations and effects provides a better understanding. In this case, the focus does not lie on the original perspective and game events, but on the player and their interaction with the game.

We can summarize the feedback of our participants into three main perceived advantages of the third-person perspective: (1) providing an entertaining experience by showing the player in action, (2) giving a good impression of the VR experience by revealing the player's full-body movement and the relation between the player's manipulations and the effects in the game world, and (3) avoiding motion sickness. Consequently, the mixed-reality third-person approach is particularly promising for games in which the player performs interesting, distinctive movements in real life.

However, the overall preference for the first-person view and the aforementioned concerns demonstrate that the third-person perspective introduces challenges that need to be taken into account. It is essential that the spectator's view is not noticeably limited: spectators must not feel that they miss important events due to a static third-person view or that the player's body might cover significant parts of the environment. Besides the higher immersive experience, these were the most prevalent reasons speaking in favor of the first-person alternative. Hence, the chosen point of view in the third-person perspective must be considered carefully. Especially in the case of dynamic viewing directions, content creators should consider integrating more dynamic solutions, such as aligning the third-person camera with the player's head rotations.

\subsection{Limitations and Future Work}
Our work presents the first step towards a better understanding of the preferences and experiences of VR game spectators with regard to the streaming perspective. While the study provides valuable insights, there are also some limitations leading to the need for further research.  

First of all, we point out that our choice of games does not represent the full VR gaming landscape. Hence, our results are limited to comparable game content. We will consider other genres (e.g., RPGs like \textit{Asgard's Wrath}~\cite{AsgardsWrath}) in the future to see which features of the two perspectives become particularly prevalent in other scenarios. Furthermore, our study does not take into account other common streaming approaches that integrate the player into the stream in other ways, such as picture-in-picture modes. While this is a limitation of our current work and should be considered in the future, our focus on the comparison of the first-person and the third-person perspectives promotes our understanding of spectator's basic preferences and desires.

Another limitation concerns the concrete implementation of the third-person perspective used in our study, in particular regarding the game \textit{Superhot VR}. As already explained in Section~\ref{sec:implementation}, the static mixed-reality approach is just one possibility to create a third-person view. 
Other possibilities include the use of a dynamic camera and the replacement of the real player footage by a virtually created avatar.
Some of the perceived shortcomings of the third-person view might trail off when using a different method, in particular a dynamic camera. For instance, the missing focus on current game events or the occlusion of relevant game objects can be reduced by automatically adapting the camera to the player's viewing direction or by enabling spectators to control the viewing angle.
While we are convinced that our choice of a static mixed-reality view is appropriate to investigate basic differences between a focus on the game (first-person) and a focus on the player (third-person), future studies with alternative implementations such as using a dynamic camera or a virtual avatar should complement and refine the findings. 

Our design decisions regarding the concrete positions of the static third-person viewpoint in the three games (i.e., the position of the spectator's camera) might have influenced the viewing experience, as well. During the design process, we experienced that some positions are better suited than others, though no position seems to be optimal in every game situation, because a static view does not adapt during the course of the game. Hence, we informally tested different positions while implementing the third-person views to find an appropriate position for each game.

Another important constituent of the third-person perspective is the player's persona.
We used the same player in all our videos to preserve comparability. Nevertheless, the specific choice introduces possible effects arising from participants' personal preferences regarding gender or appearance of the player. Hence, future studies should include other types of players to investigate potential impacts.
Furthermore, preferences may change if the viewers have some kind of relationship with the content producer, for instance, if the player is their favorite streamer. One participant explicitly stated: \textit{"If it's my usual go-to streamer on Twitch, then I would probably like the third-person perspective better because it'd be funnier."}. In such cases, the focus of interest is more on the player and less on the game, which is much better supported by a mixed-reality perspective.

Previous research also indicates that the social interaction between the player and the audience can be an important motivator for spectators to follow game streamers~\cite{hamilton2014streaming, sjoblom2017people}. Particularly in a live streaming context, viewers' social motives become more prevalent, as they have the possibility to interact with the streamer or other viewers while the action takes place.
In our study, we presented prerecorded videos and decided to not include social features (i.e., the player did not speak to the audience) to reduce the potential interference effects caused by our specific player. This approach increased the controllability of the study procedure, but limits the direct transferability of our results to live streaming.
We assume that the pros and cons of the different perspectives found in our study also apply to live streaming contexts and that our results can also inform live streaming design choices. For instance, streamers using a first-person perspective can further increase comprehensibility by verbally describing which movements they are performing (because these are not visible to the audience).

However, our study does not provide direct indications on how the different perspectives support or interfere with the viewers' need for social interaction. It might become more important to see the player, as visual cues are a central aspect in human communication. On the other hand, a first-person view might provide a closer connection to the player, because this perspective fosters a shared focus and attentional allocation. Future research is needed to test such assumptions. Hence, as a complement to our current work, we recommend the conduct of in-the-wild studies on streaming platforms with actual streamers and their audiences to capture this important social aspect with regard to the preferences of different perspectives. This would also enable a more sophisticated investigation of the correlations between spectators' motives to watch a certain VR game stream and their preferred view.

Another interesting direction for future research in the area of VR spectatorship would be to investigate the experience of spectating VR game streams using HMDs. If the viewer is equipped with an immersive HMD, there are different possibilities to present VR content and the viewing experience will probably differ from 2D displays.

\section{Conclusion} 
Delivering the highly immersive experience of VR games to a broad audience via common 2D video streams is a challenge for VR content providers, such as streamers, advertisers, and game developers. This work offers support by giving advice on the choice of an appropriate spectator perspective to foster a positive viewing experience. 

Based on our study results, we identified two key factors that need to be considered when deciding between a first-person and a mixed-reality third-person perspective: first, the characteristics of the game, in particular the location of game events in relation to the player's position; and second, the motives and expectations of the audience. While the first-person perspective puts the focus on the game and resembles the player's view, the mixed-reality third-person perspective shifts the focus to the player and the player-game interaction. For games in which most game events evolve directly around the player, the third-person perspective provides viewers with unique insights by revealing the player's real movements and their effects in the game world. This positive effect of the third-person perspective particularly applies to games that require the player to perform interesting, distinctive movements. 
In contrast, if the main game action is distributed over the game environment and not centered around the player, a first-person perspective is more appropriate due to its immersive quality and a clear focus on relevant game events. 

Apart from the game characteristics, content providers also should consider their audience. If spectators are supposed to be mainly interested in gaining an impression of the game and less in the player's persona, the first-person view provides the desired information better than the third-person perspective. On the other hand, if spectators have a keen interest in a specific streamer, their preference might be biased towards a third-person view, which highlights this person. 

This work presents the first step towards a comprehensive VR streaming guideline. As discussed above, there are some limitations and follow-up studies with more VR games and different settings are needed to extent our current knowledge. 
In particular, other implementations of the third-person perspective, for instance with a dynamic camera, need to be investigated to test our hypothesis about the importance of player centricity and visual coupling.
Our work paves the way for further research on the spectators' experiences and expectations in the context of VR content.
In the long term, understanding how different perspectives contribute to different demands of VR spectators will foster informed design decisions in diverse application areas such as game streaming, VR training and mixed-reality multi-user scenarios.

\balance

%% The next two lines define the bibliography style to be used, and
%% the bibliography file.
\bibliographystyle{ACM-Reference-Format}
\bibliography{references}

\end{document}